\documentclass[prd,aps,twocolumn,superscriptaddress,longbibliography,nofootinbib]{revtex4-1}

\usepackage{amsmath,graphicx,epsfig}
\usepackage{epstopdf}
\usepackage{euscript}
\usepackage{amsfonts}
\usepackage{amssymb}
\usepackage{float}
\usepackage{tabularx}
\usepackage{amsmath,amsfonts}
\usepackage{soul,xcolor}
\usepackage{mathtools}

\DeclarePairedDelimiter\ket{\lvert}{\rangle}
\DeclarePairedDelimiterX\braket[2]{\langle}{\rangle}{#1 \delimsize\vert #2}
\usepackage{physics}
\begin{document}
\draft
\title{\bf Oscillating nuclear electric dipole moments inside atoms}
\author{Victor~V.~Flambaum} 
\affiliation{School of Physics, University of New South Wales,  Sydney 2052,  Australia}
\affiliation{Helmholtz Institut, Johannes Gutenberg-Universit{\"a}t Mainz, 55128 Mainz, Germany}
\affiliation{The New Zealand Institute for Advanced Study, Massey University Auckland, 0632 Auckland, New Zealand}
\author{Dmitry Budker}
\affiliation{Helmholtz Institut, Johannes Gutenberg-Universit{\"a}t Mainz, 55128 Mainz, Germany}
\affiliation{Department of Physics, University of California, Berkeley, California 94720, USA}
\author{Arne Wickenbrock}
\affiliation{Helmholtz Institut, Johannes Gutenberg-Universit{\"a}t Mainz, 55128 Mainz, Germany}
\date{\today}
\begin{abstract}
Interaction with the axion dark matter (DM) field generates an oscillating nuclear electric dipole moment (EDM) with a frequency corresponding to the axion's Compton frequency. Within an atom, an oscillating EDM can drive electric dipole transitions in the electronic shell. In the absence of radiation, and if the axion frequency matches a dipole transition, it can promote the electron into the excited state. The excitation events can be detected, for example, via subsequent fluorescence or photoionization.
Here we calculate the rates of such transitions. For a single light atom and an axion Compton frequency resonant with a transition energy corresponding to 1\,eV, the rate is on the order of 10$^{-22}$ per year, so a macroscopic atomic sample would be needed. A fundamental challenge is discriminating against background processes that may lead to the excitation of the same electric dipole transition.  
The ways to enhance the signals to potentially observable levels exceeding backgrounds and to search for axions in an extended frequency range are discussed.
\end{abstract}
\maketitle
{\bf Introduction}. The nature of dark matter remains unknown.
The axion is a prominent dark matter candidate originally introduced in the 1970s to explain the apparent charge-parity (CP) symmetry of the strong interactions. Most searches for axion and axion-like particles (ALPs\footnote{ALPs are pseudoscalar particles like the axion that do not, however, solve the strong-CP problem; we refer to both axions and ALPs as ``axions'' in this paper.}) rely on conversion between axions and photons. Recently, experiments like the Cosmic Axion Spin Precession experiments (CASPEr) started to look for other types of axion couplings \cite{Casper}. Assuming the dark matter in the Milky Way consists of axions of a given sub-eV mass, the dark matter can be described as a field oscillating at the axion Compton frequency. This field induces oscillating electric dipole moments (EDM) of fundamental particles, nuclei, atoms, and molecules \cite{Stadnik2014,Graham} and causes precession of particle's spins due to gradients in the axion field (the axion-wind effect) \cite{GrahamRajendran}. The CASPEr experiments search for spin precession due to axion-induced EDM and axion wind with nuclear magnetic resonance. First results constraining the axion-nucleon couplings have been published by CASPEr \cite{Antoine, Comagnetometer}, as well as by other groups re-analyzing existing data obtained in the neutron-EDM \cite{nEDM} and atomic co-magnetometer experiments \cite{VolanskyArxiv}.

In this note, we analyze the effect of an axion-induced oscillating EDM
in an atom. Such a dipole moment may induce an electric-dipole (E1) transition in the atom if the transition frequency matches the axion Compton frequency. We present an estimate of the corresponding transition rates, discuss their scaling with the relevant atomic parameters and comment on the feasibility of experimental observation of the effect.

{\bf Nuclear EDM produced by the axion field}.  As noted in Ref.\,\cite{Witten} that a nucleon EDM may be produced by the so-called ``QCD $\theta$-term.'' Numerous references  and recent results for the neutron and proton EDM  are summarised in the review \cite{Yamanaka}:
 \begin{eqnarray}\label{theta}
 d_n= -(2.7 \pm 1.2) \times 10^{-16} \theta \, e\,  \textrm{cm} \,,\\
 d_p= (2.1 \pm 1.2) \times 10^{-16} \theta \, e\,  \textrm{cm}\,.
\end{eqnarray}
Calculations of the nuclear EDM produced by the P,T-odd nuclear forces were performed in Refs.\,\cite{HH,SFK,FKS1985,FKS1986}. For a general estimate of the nuclear EDM it is convenient to use a single-valence-nucleon formula from \cite{SFK} and express the result in terms of $\theta$ following Ref. \cite{FDK}:
\begin{equation}\label{d}
   d \approx   7 \times 10^{-16}  \left(q- \frac{Z}{A}\right)\left(1- 2q\right)  \left<\sigma\right> \theta \, e  \, \textrm{cm}.
  \end{equation}  
  Here $q=1$ for a valence proton, $q=0$ for the valence neutron, the nuclear spin matrix element is  $\left<\sigma\right>$ = 1 for a valence nucleon with $j=l+1/2$ and $\left<\sigma\right>$ = -j/(j+1) for a valence nucleon with $j=l-1/2$, where $j$ and $l$ are the total and orbital angular momenta of the valence nucleon.

There are many specific results for the  $^2$H and $^3$He EDM, see e.g. reviews \cite{Yamanaka,Chupp}. Within the error bars the deuterium EDM is consistent with zero due to the cancellation between the proton and neutron contributions. The $^3$He nucleus contains an unpaired neutron. Using the calculation of the contribution of the T,P-violating nuclear forces from Ref.\,\cite{DeVries} ($-1.5 \times 10^{-16} \theta \, e\,  \textrm{cm}$) and the value of the neutron  EDM  from Eq.\, \eqref{theta} we obtain for the $^3$He EDM:
\begin{equation}\label{3He}
   d(^3\textrm{He}) = \left(- 4.2 \pm 1.5\right)  \times 10^{-16}  \theta \, e  \, \textrm{cm}.
  \end{equation}  
This can be compared with an estimate using Eq.\,\eqref{d} which gives $d(^3\textrm{He}) = - 4.7  \times 10^{-16}  \theta \, e  \, \textrm{cm}$.

Reference \cite{Graham} discussed the possibility that the dark matter field is, in fact, an oscillating $\theta$-term corresponding to an axion field that generates an oscillating nucleon EDM. Relating the value of the axion field to the local dark matter density (Ref.\,\cite{Stadnik2014}) we may substitute $\theta(t)=\theta _0 \cos(\omega t)$ where $\theta _0\approx 4 \times 10^{-18}$, $\omega=m_a c^2/\hbar$ and $m_a$ is the axion mass. It is important to keep in mind that ALPs inducing larger dipole moments are also among viable DM candidates, so an experiment with sensitivity less than that necessary to detect axion DM could already be sensitive to DM composed of such ALPs.

\color{black}{\bf Oscillating nuclear EDM induced transitions.} The term in the Hamiltonian of an atom accounting for the interaction of the atomic electrons with the field of an oscillating nuclear EDM $\bf d$ may be presented as 
\begin{equation}\label{Phi}
V= 
e\sum_{k=1}^{N}\frac{{\bf d r}_k}{r_k^3}=\frac{i}{Z e \hbar}[{\bf P},H_0]\bf d \,,
\end{equation}
where $H_0$ is the Schr{\"o}dinger or the Dirac Hamiltonian for the atomic electrons in the absence of an oscillating EDM, $N$ is the number of the electrons, $Z$ is the nuclear charge, 
$- e$ is the electron charge, $r_k$ is  the electron position relative to the nucleus, and
\begin{equation}\label{P}
{\bf P}=\sum_{k=1}^{N}{\bf p}_k
\end{equation}
is the total momentum of all atomic electrons, which commutes with the electron-electron interaction but does not commute with $U$, the potential energy due to the interaction with the nucleus.
\begin{equation}\label{U}
U=-\sum_{k=1}^{N}Ze^2/r_k\,.
\end{equation}
The remaining commutator can be written as:
\begin{equation}\label{Commutator}
[{\bf P },H_0] = [{\bf P }, U] = -i \hbar  Z e^2 \sum_{k=1}^{N}\nabla \frac{1}{r_k}\,.
\end{equation}
We assume that the nuclear mass is infinite and neglect the Breit and magnetic interactions. Using  $H_0 \ket{n} = E_n \ket{n}$  we obtain the matrix element of $V$ between atomic states $\ket{1}$ and $\ket{2}$: 
\begin{equation}\label{V12}
\mel{2}{V}{1}=-
\frac{i\omega_{21}}{Ze} \mel{2}{\bf P}{1}\bf d\,,
\end{equation}
where $\omega_{21}=(E_2-E_1)/\hbar$.  
With the non-relativistic relation  
\begin{equation}\label{V13}
{\bf P}=- \frac{i m} {e \hbar} [H_0,{\bf D}]\,,
\end{equation}
where $m$ is the electron mass and  
\begin{equation}\label{V14}
{\bf D}=-e \sum_{k=1}^{N} {\bf r}_k\,,
\end{equation}
is the electronic dipole operator.  We express the result in terms of the electric-dipole transition amplitude $E1=\mel{2}{D}{1}$: 
\begin{equation}\label{VE1}
\mel{2}{V}{1}=\frac{\omega_{21}^2 m}{Ze^2}\mel{2}{\bf D }{1}\bf d\,.
\end{equation}
The scalar operator $V$ conserves the total atomic angular momentum $F$. The selection rules for the electron variables  are identical to those for E1 transitions. 

The matrix element of Eq.\,\eqref{VE1} is proportional to the square of the transition frequency so it vanishes in the limit of low frequencies. As discussed below, this is related to the Schiff theorem stating that a static EDM of a subatomic point particle is unobservable in the nonrelativistic approximation. 

The transition probability $W \propto  \mel{2}{V}{1}^2 $ is inversely proportional to the nuclear charge squared,  $W \propto 1/ Z^2$, i.e. light atoms like H, He, Li have are advantageous for experiments. The origin of the factor can be related to the Schiff theorem: the screening factor for an external electric field scales $\propto Z$.

Note that the transition probability is not suppressed for high electron waves $j,l$. The reason is that the  matrix element of $V \propto 1/r^2$ does not converge at small distances. Indeed, an estimate  integral $\int (\psi_1^+\psi_2 /r^2) r^2 dr = 
 \int \psi_1^+\psi_2  dr  \propto r^{l_1+l_2 +1}$ converges at the atomic size. This is also the reason why the relativistic corrections are not important (except for the values of the energies $E_1,E_2$).  
 The matrix element rapidly decreases with the electron principal quantum number $n_{1,2}$ since $\omega_{12} \propto (n_2-n_1)/n_{1,2}^3$, $\mel{2}{D}{1} \propto n_{1,2}^2$, i.e. $\mel{2}{V}{1} \propto 1/n_{1,2}^4$. Here we assumed $n_1\approx n_2\approx n_{1,2}$. \\
{\bf Transition probability}. The probability of the transition on resonance for the stationary case (time $t \gg 1/\Gamma$) for the perturbation $V=V^0 \cos{\omega t} $ is given by the following expression \cite{Landau}:
\begin{equation}\label{W01}
W_{01}=\frac{|\mel{2}{V^0}{1}|^2}{\hbar^2 \Gamma} t \,.
\end{equation}
Here $\Gamma$ is the width of the virialized axion frequency distribution ($\Gamma \approx 10^{-6} m_a c^2/\hbar =10^{-6} \omega$) if the atomic transition width is smaller than the axion energy distribution width (or the atomic transition width vice versa). Inserting Eq.\,\eqref{VE1} into \eqref{W01} and substituting the above values,  we obtain the approximate time for one transition:
\begin{align}
\label{Eq: t} 
&t= \nonumber\\
&\frac{2\,10^{22}}{N} Z^2\left(\frac{1 \text{\,eV}}{\hbar \omega}\right)^3 \left(\frac{3 e a}{\left|\mel{2}{D_z}{1}\right |}\right)^2 \left(\frac{ 4\,10^{-16} \theta \, e\,  \textrm{cm}}{d}\right)^2 \, \textrm{y},
\end{align}
where $N$ is the number of atoms and  $a$ is the Bohr radius. We presented the result for the maximal projection of the atomic angular momentum ($F_z=F=J+I$, where $J$ is the electron angular momentum and $I$ is the nuclear spin) and normalised the result to a transition frequency corresponding to 1\,eV, a typical value for low-lying atomic states of an allowed E1-transition amplitude $|\mel{2}{D_z}{1}|=3 ea$  and a typical nuclear EDM $d=4 \times 10^{-16} \theta \, e\,  \textrm{cm}$. It is easy to obtain specific values for the hydrogen atom and transitions between highly excited Rydberg states of electrons where there are analytical expressions \cite{Bethe} for the transition frequencies,  $\omega_{21} \approx$13.6\,eV$ (n_2-n_1)/n_{1,2}^3$ and E1 amplitudes $\mel{2}{D_z}{1}\approx n_{1,2}^2 a$. Altogether we have $t \propto Z^2 n_{1,2}^{5}$ further informing the choice of the transitions.

{\bf Discussion and conclusion}
In this note, we have asked and answered the question: can axion-induced nuclear EDM drive an atomic transition? The answer is yes, albeit the rate of the transitions is low. While the search would be limited to axion frequencies closely corresponding to resonant atomic transitions, the latter can be tuned by using Zeeman and Stark effects. In addition, the transitions are ``dense'' in the region of Rydberg excitations, so  complete coverage of a frequency interval is, in principle, possible (although the rate is suppressed as $\propto n^{-5}$, see Eq. \eqref{Eq: t}. Molecules also have dense spectra and may present additional advantages for this kind of experiments \cite{flambaum2019oscillating}.

As mentioned above, the rate per atom is ``astronomically'' low [see Eq.\,\eqref{Eq: t}]. In principle, one could counteract this smallness by utilizing a large sample of, say $\approx 10^{30}$ atoms corresponding to several tons of material. The rates can also be enhanced by interfering the EDM-induced transition amplitude with a reference amplitude induced by applying a resonant electromagnetic field. In this case, the signal is linear (rather than quadratic) in the weak coupling but the interference term will randomly fluctuate with the coherence time of the axion.  A more fundamental problem is how to distinguish the axion-induced excitation from a variety of possible background processes, e.g., excitation by black-body radiation. In this respect it is particularly ``unhelpful'' that the transitions are of E1 type and there do not seem to be any selection rules to distinguish the sought-for events. We note, that it is just such selection rules that enable many atomic symmetry-test experiments like the ones that measure parity violation \cite{Safronova2018}.

The oscillating EDM induced transition rate vanishes in the static limit [$\omega\rightarrow 0$; Eq.\,\eqref{Eq: t}]. This result can be related to the Schiff theorem that can be formulated to state that a static nuclear EDM does not induce an atomic EDM. However, this suppression should not appear for the transitions induced by the oscillating Schiff and magnetic quadrupole moments which can also be induced by the axion dark matter field.  However, here there is a suppression due to a small nuclear size $R_N$  [for the Schiff-moment contribution, the suppression goes as $R_N^2 /(a/Z)^2$], which is reduced by the large $Z^2$ in heavy atoms.

While in the present note we consider the effect of an axion-induced nuclear EDM, other couplings may have different manifestation and lead to excitation of different transition types. For instance, the axion's derivative coupling to nucleons or electrons results in an oscillating magnetic moment that can drive magnetic-dipole (M1) transitions. 

Finally, the source of axions does not necessarily be the galactic dark matter. Axions may also be produced in the sun or via conversion from laser photons. In this case, they can be of relatively high frequency (even if their mass is low) relaxing the $\omega^3$ suppression in Eq.\,\eqref{Eq: t}.   
\color{black}
{\bf Acknowledgment}. The authors thank Mikhail~G.~Kozlov, Derek~F.~Jackson Kimball and Alexander~O.~Sushkov for helpful discussions. This work is supported by the Australian Research Council, a Gutenberg Fellowship, and New Zealand Institute for Advanced Study. It has also received support from the European Research Council (ERC) under the European Unions Horizon 2020 research and innovation program (grant agreement No 695405), from the DFG Reinhart Koselleck project and the Heising-Simons Foundation.
\bibliography{bib}
\end{document}